\documentclass{article}

\usepackage{arxiv}

\usepackage[utf8]{inputenc} 
\usepackage[T1]{fontenc}    
\usepackage{hyperref}       
\usepackage{url}            
\usepackage{booktabs}       
\usepackage{amsfonts}       
\usepackage{nicefrac}       
\usepackage{microtype}      
\usepackage{lipsum}		
\usepackage{graphicx}
\usepackage{natbib}
\usepackage{doi}

\title{Low-Noise Operation of Stepped Frequency-Comb Sources Based on Phase-Code Mode-Locking}

\author{ Tae-Shik~Kim \\
	Wellman Center for Photomedicine\\
	Massachusetts General Hospital\\
        Harvard Medical School\\
	Boston, MA 02114 \\
	\And
        \href{https://orcid.org/0000-0003-2768-3684}{\includegraphics[scale=0.06]{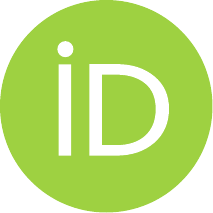}\hspace{1mm}Danielle J.~Harper}\\
        Wellman Center for Photomedicine\\
        Massachusetts General Hospital\\
        Harvard Medical School\\
        Boston, MA 02114\\
        \And
	\href{https://orcid.org/0000-0002-6623-3515}{\includegraphics[scale=0.06]{orcid.pdf}\hspace{1mm}Benjamin J.~Vakoc}\\
	Wellman Center for Photomedicine\\
        Massachusetts General Hospital\\
        Harvard Medical School\\
        Boston, MA 02114\\
	\texttt{bvakoc@mgh.harvard.edu} \\
}

\date{}

\renewcommand{\shorttitle}{\textit{arXiv} Template}

\begin{document}
\maketitle

\begin{abstract}
The phase-code mode-locked (PCML) laser provides a configurable frequency comb light source for circular-ranging optical coherence tomography systems (CR-OCT). However, prior implementations of PCML suffered from high relative intensity noise (RIN). In this work, we demonstrate that by configuring the output pulsewidth and pulse separation in relation to the linewidth of the intracavity Fabry-Perot etalon, low-noise operation can be achieved. We observed a more than ten-fold reduction in laser RIN that enabled CR-OCT imaging with a sensitivity of 103 dB with 35 mW delivered to the sample and provided images that are qualitatively comparable to those acquired using traditional swept-source methods. 
\end{abstract}

\section{Introduction}
Conventional methods in optical coherence tomography (OCT) generate images in which there is a straightforward one-to-one mapping between the image space and the physical space\cite{izatt2008theory}. In contrast, circular-range (CR) OCT methods create images with a one-to-many mapping. A single location in the generated CR-OCT image describes the summed reflections of multiple, equally spaced physical delays \cite{siddiqui2012optical,siddiqui2018high,bouma2022primer}. In some settings, this can enable an efficient capture of the sample's local structure without \textit{ a priori} knowledge of the sample location. 

Conventional OCT and CR-OCT systems are differentiated by their light sources. Whereas conventional OCT uses a continuously broadband optical source, CR-OCT uses stepped frequency-comb sources that output each combline in sequence, resulting in a train of nearly single-frequency optical pulses. The stepped frequency-comb source allows the sample response at each combline frequency to be independently measured using high-speed photodetectors. Several stepped frequency comb laser architectures have been described \cite{tozburun2014rapid, lippok2019extended, khazaeinezhad2017spml}, including one based on phase-code mode locking (PCML) \cite{kim2020stepped}. In PCML lasers, a pair of electro-optic phase modulators are combined with a fixed Fabry-Perot (FP) etalon and intracavity chromatic dispersion to enable electronically controllable stepped frequency comb outputs. PCML is highly configurable (and reconfigurable) and straightforward to synchronize with data acquisition systems. However, prior PCML demonstrations suffered from high relative intensity noise (RIN) that limited imaging sensitivity.  

In this work, we demonstrate a second-generation PCML source that reduces RIN noise by a factor of ten, achieving noise performance comparable to or better than conventional OCT sources. This is done without adding any additional elements to the source. Instead, we identify regimes of low-noise operation defined by the relationship of the output pulsewidth and output pulse separation to the characteristics of the intracavity FP etalon. We demonstrate improved OCT imaging using the low-noise PCML source.  

\section{Theory, Methods and Materials}

\subsection{PCML operating principle}

\begin{figure}[tb]
\centering\includegraphics[width=8.5cm]{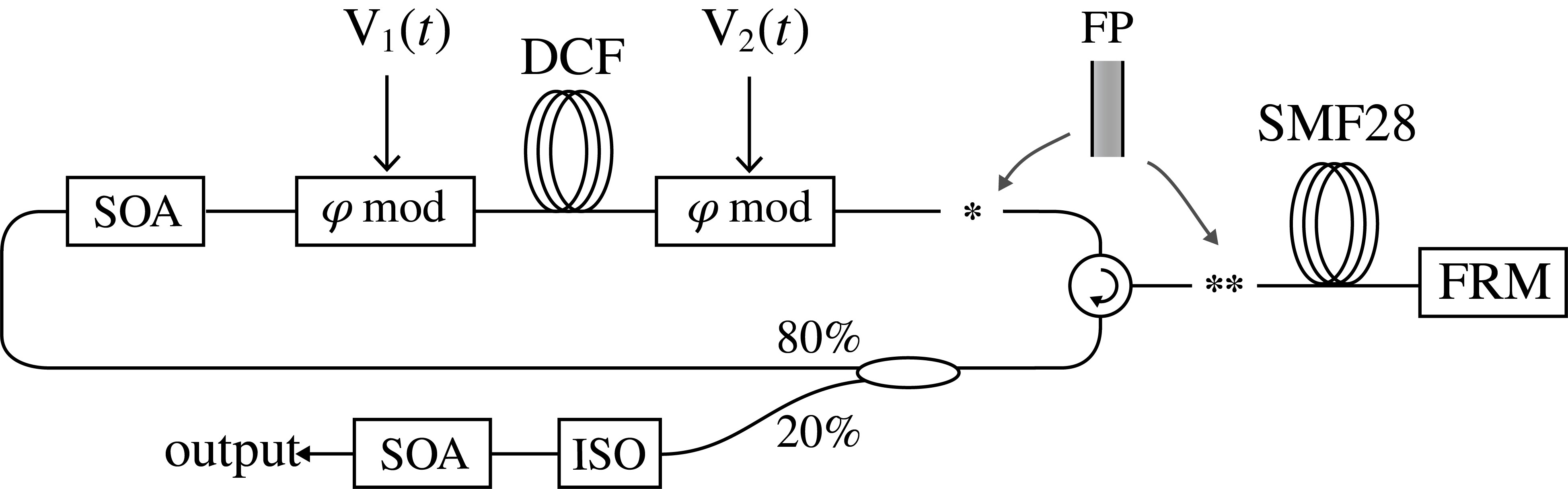}
\caption{The PCML laser architecture including a single-pass (*) and double-pass (**) location for the Fabry-Perot (FP) etalon. SOA: semiconductor optical amplifier. $\phi$ mod: lithium-niobate phase modulator. DCF: dispersion compensating fiber. SMF28: Corning single-mode fiber. FRM: Faraday rotator mirror. ISO: optical isolator.}
\label{fig:PCML}
\end{figure}

We first review the architecture and operating principle of the PCML laser, building upon the original demonstration in~\cite{kim2020stepped}. The architecture of the PCML laser includes an intracavity fixed FP etalon that defines the free-spectral range (FSR) of the frequency comb, a semiconductor optical amplifier (SOA), a pair of electro-optic phase modulators, and matched positive and negative dispersive fibers. The architecture of the PCML laser used in this work (Fig. \ref{fig:PCML}) is similar to that of our previous work \cite{kim2020stepped}. 

The PCML laser is based on the reversibility of spectral broadening induced by rapid phase modulation. In the PCML, a first electro-optic modulator serves to phase modulate any combline output from the FP etalon, resulting in a spectral broadening on each wavelength of the comb. We refer to this as phase-encoding. The second phase modulator then acts on these optical signals and is designed to provide the opposite phase modulating waveform as that provided by the first modulator. This is phase-decoding, and, if timed correctly, exactly reverses the spectral broadening induced by the first phase modulator. Because of the inclusion of chromatic dispersion between the modulators, each optical combline signal experiences a different travel time from the first to the second. As such, the second phase modulator is timed to reverse the spectral broadening for only one combline at a time. By shifting the delay of the drive signal provided to the second phase modulator, different optical comblines can be selected for phase "decoding," i.e., spectral re-narrowing. Now, these optical signals travel to the FP etalon, which passes the decoded (narrowed) combline with lower loss, providing favorable conditions for lasing at that combline. The comblines that are not correctly decoded experience either an incomplete spectral narrowing (if the decoding waveform was relatively close to the opposite of the encoding waveform) or a further spectral broadening (if the decoding waveform is more fully uncorrelated to the encoding waveform).  

To use this principle in a frequency-agile laser, the delay of the waveform provided to the second (decoding) phase modulator is varied over time. To allow the output to change more rapidly than the cavity round-trip time, this delay is varied synchronously with the cavity round-trip time, following mode-locking principles such as those used in FDML\cite{huber2006fdml} and SPML lasers\cite{tozburun2014rapid}\cite{khazaeinezhad2017spml}. 

Although the operating principle of the PCML is somewhat complex, it can be implemented with a relatively straightforward optical design and a two-channel high-speed arbitrary waveform generator (AWG). We used an AWG (AWG872, Euvis, Newbury Park, CA) that provided two channels at 8 GSPS. These waveforms can be used to define the sequence of optical frequency outputs, but also to define the temporal width of each optical frequency pulse and, in addition, the off-time between these pulses. This work focuses on the role that these timing parameters play on the noise performance. In the next section, we define the relevant timing parameters for a PCML source and then describe the AWG signal construction used to realize these specific timed source outputs. 

\subsection{FP response time}

In the agile PCML laser, the drive signals provided to the phase modulators control the output pulse stream. This includes which optical frequency is selected within a given pulse, and also the duration of each pulse and the temporal separation between pulses. The central insight of this work is the recognition that the intracavity FP etalon sets an intrinsic time-scale, and that by aligning the output pulse timing to this intrinsic time-scale, the PCML stability and noise are greatly improved. 

\begin{figure}[]
\centering\includegraphics[width=12.0cm]{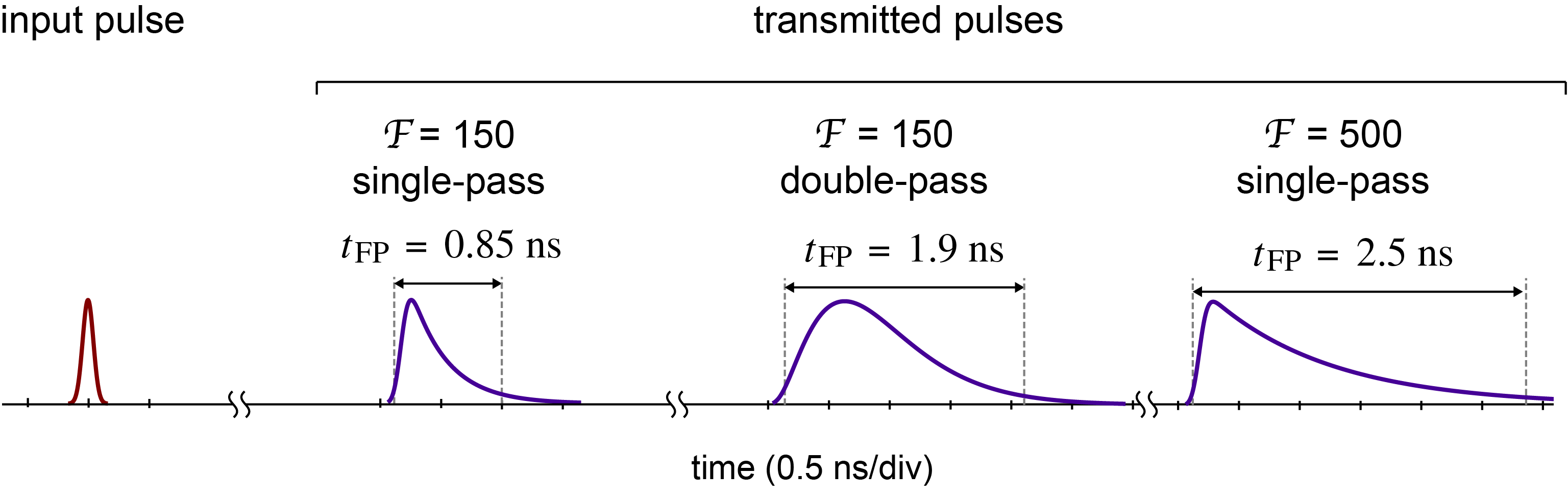}
\caption{The calculated impulse response for three etalon configurations used in this paper. (a) Single-pass; Finesse = 150. (b) Double-pass; Finesse = 150. (c) Single-pass; Finesse = 500. Not illustrated is the etalon impulse response for a Finesse = 8000 etalon operated in single pass configuration ($t_{\textrm{FP}}$ = 36 ns).}
\label{fig:Impulse_Response}
\end{figure}

The FP etalon creates a periodic transmission (amplitude) filter within the laser cavity. This etalon must also, by causality, modulate the phase response of an input optical field. Consequently, a light pulse passing through the etalon will be delayed and temporally smeared, meaning that there is an inherent effective response time associated with a given FP etalon. This is illustrated in Fig.~\ref{fig:Impulse_Response} for three of the four experimental configurations of this paper: single-pass of an etalon with a Finesse = 150 (Fig.~\ref{fig:Impulse_Response}(a)); double-pass with the same Finesse = 150 etalon, approximately increasing the Finesse to 300 (Fig.~\ref{fig:Impulse_Response}(b)); and single-pass of a Finesse = 500 etalon (Fig.~\ref{fig:Impulse_Response}(c)). For each, the output pulse is shown for a short input pulse with a center optical frequency aligned to a transmission peak of the etalon. This can be considered an impulse response for the etalon. As expected, as the linewidth of the etalon decreases (through increasing Finesse), the response time of the etalon increases. Figure~\ref{fig:PCML} shows the position of the FP etalon in both single-pass and double-pass configurations for the experiments conducted in this work.

The FP etalon response time is often quantified through the etalon lifetime, or etalon ring-down time/decay time. However, this parameter is not easily calculated based on the etalon properties for a double-pass configuration such as that shown in Fig. \ref{fig:Impulse_Response}(b) \cite{vaughan2017fabry}. Instead, we will quantify the etalon response time as the width of the etalon's impulse response at 10\% of the peak height, and using a Gaussian input pulse that is 1) much less than the response time and 2) aligned in optical frequency to the etalon transmission peak. This impulse response can be calculated by summation of the circulating fields. For the etalons used in this work, the response times, $t_{\textrm{FP}}$, are 0.85 ns, 1.9 ns, and 2.5 ns for the three configurations illustrated in Fig.~\ref{fig:Impulse_Response}, and 36 ns for a fourth Finesse = 8000 etalon (not shown in Fig.~\ref{fig:Impulse_Response}). Note that the shape of the impulse response for the double-pass configuration differs from that of the single-pass configuration; the rise-time is extended.

\subsection{Phase-Modulator Drive Signal Construction}

\subsubsection{Linewidth broadening}

The waveform sent to the first phase modulator broadens the linewidth of all wavelengths of the frequency comb input equally. For stable PCML laser generation, this waveform should be non-periodic, i.e., lacking any dominant frequency component in the Fourier domain. In order for each pulse to contain exactly one optical combline, there should be only one time delay between broadening and compensation, which results in lasing at any given time. Periodicity would imply that there would be more than one time delay that could provide broadening compensation, increasing the risk of unwanted multi-wavelength excitation within the optical bandwidth. A chirped sinusoid meets this criteria and was selected for this work.

The frequency range of the chirped sinusoid is important. If the frequency range is too low, it starts to approximate the periodicity of a regular sinusoid, sacrificing the extinction between adjacent lines. At the other end of the scale, the maximum frequency range is dictated by the sampling rate of the arbitrary waveform generator (AWG) used to generate the waveforms. Since the AWG sample rate is finite, a high frequency range will result in a loss of precision, thereby causing loss through the phase code filter. With these considerations in mind, we selected a chirped sinusoid with frequencies ranging from 900 MHz - 1700 MHz, appropriate for an AWG with a sampling rate of 7.4 GSPS. The duration of a single sweep of this chirped sinusoid, $t_{\textrm{ON}}$, is the programmed on-time, or intended pulsewidth, of the laser.   

A key improvement of this work over that of our previous PCML design \cite{kim2020stepped} is that of the addition of "off-time" in between the PCML pulses. This is necessary for the stability of the PCML and the reduction of relative intensity noise (RIN). A sinusoidal wave with a frequency of 2.2 GHz was sent to the first phase modulator during this time, providing enough RF power to broaden the pulse but at a frequency that cannot be recovered by the inverse chirp during linewidth recovery. The duration of this sinusoidal signal,  $t_{\textrm{OFF}}$ is the programmed off-time, or separation between pulses, of the source.

\subsubsection{Linewidth recovery}

The corresponding linewidth recovery signals consisted simply of the inverse of the chirped sinusoid, delayed with respect to the start of the first phase modulator's waveform by fixed delay times, $\tau$, corresponding to the time it takes for each wavelength to pass the dispersive medium and reach the second phase modulator. As for the first modulator, a sinusoidal wave was sent to the second during the "off" time, this time with a frequency of 2.8 GHz. The addition of the second "off" tone ensured the PCML went off completely.

\subsubsection{Timing metrics}
\begin{figure}[]
\centering\includegraphics[width=8.5cm]{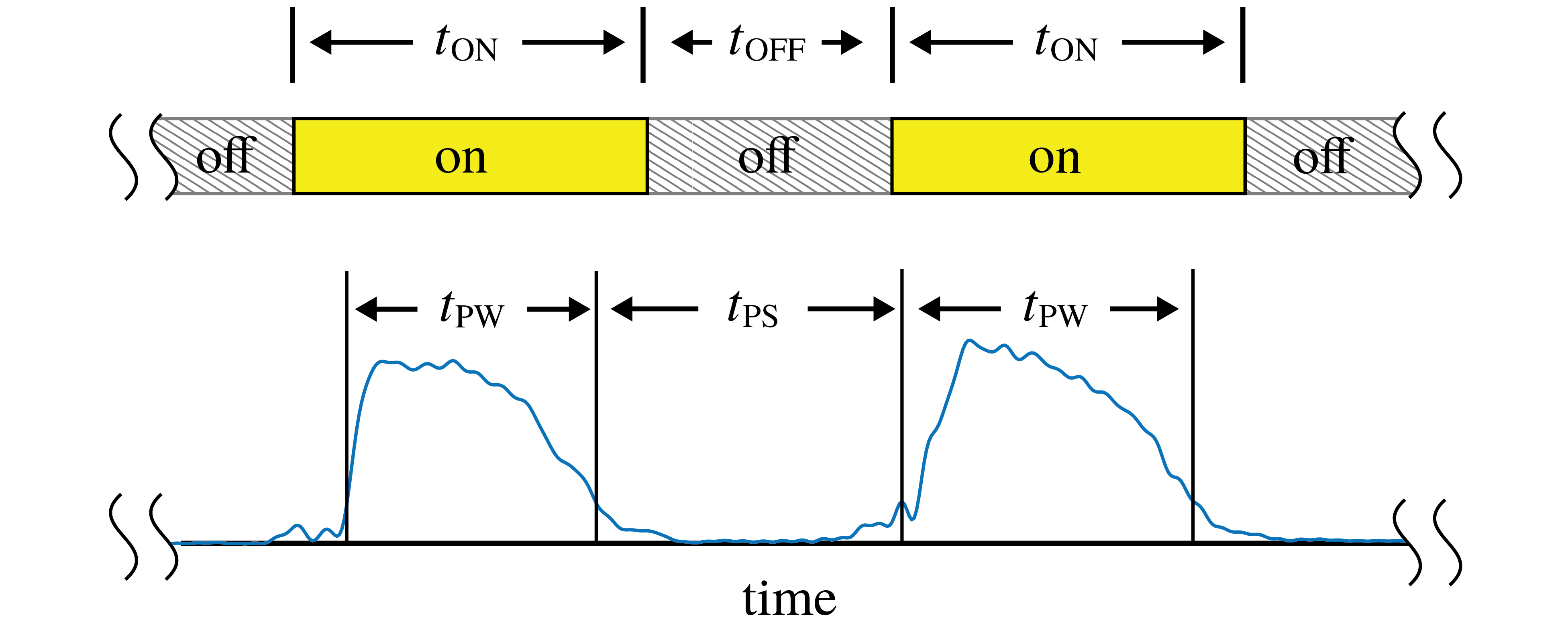}
\caption{An illustration of the PCML output timing parameters of pulsewidth and pulse separation. Each of these parameters can be defined as measured (pulsewidth: $t_\textrm{PW}$, and pulse separation: $t_\textrm{PS}$) or as commanded (through the drive waveforms, $t_{\textrm{ON}}$ and $t_\textrm{OFF}$). In this work, we describe the laser behavior based on  $t_\textrm{PW}$ and $t_\textrm{OFF}$, which appear to be the most predictive of noise. Note that the horizontal time axis in this figure is illustrative and does not imply an exact $t_{\textrm{ON}}$ to $t_\textrm{PW}$ relationship.}
\label{fig:PCML_logic}
\end{figure}

Practically, the programmed $t_{\textrm{ON}}$ and $t_{\textrm{OFF}}$ correpond well with, but are not identical to, the actual measured pulsewidth and separation. We therefore also define the corresponding empirical metrics of $t_{\textrm{PW}}$ and $t_{\textrm{PS}}$, which refer to the measured pulse width and pulse separation, respectively. These relationships are illustrated in Fig.~\ref{fig:PCML_logic}. In the following results, we will use $t_{\textrm{PW}}$ and $t_{\textrm{OFF}}$ to define stability regimes simply because these appear to be more predictive of noise. The broader observation of stability regimes holds regardless of which specific metric is selected. 

\subsection{Experimental configuration}

We operated the PCML laser with 4 separate FP etalon configurations: a single-pass etalon with Finesse of 150 yielding $t_{\textrm{FP}}$ = 0.85 ns (Fig.~\ref{fig:Impulse_Response}(a)); a single-pass etalon with Finesse of 300 yielding $t_{\textrm{FP}}$ = 2.5 ns (Fig.~\ref{fig:Impulse_Response}(c)); a single-pass etalon with Finesse of 8000 yielding $t_{\textrm{FP}}$ = 36 ns (not pictured); and finally a double-passing of the etalon with Finesse of 150 yielding $t_{\textrm{FP}}$ = 1.9 ns (Fig.~\ref{fig:Impulse_Response}(b)). The FSRs of all three etalons were between 80-85 GHz.

\subsubsection{Investigating the relationship between $t_{\textrm{PW}}$ and $t_{\textrm{FP}}$}

To investigate the relationship of $t_{\textrm{PW}}$ and $t_{\textrm{FP}}$, we programmed the PCML laser to create a sequence of 100 pulses with gradually increasing pulsewidths. The pulsewidths ranged from 0.134 ns - 13.4 ns (except for the Finesse = 8000 configuration, which was 10$\times$ longer) and spanned the cavity round-trip time of the laser. A large pulse-separation was included to isolate the role of $t_{\textrm{PW}}$ on output noise (it is shown later in this work that noise is lowest for large $t_{\textrm{OFF}}$). In effect, we programmed a PCML "A-line" wherein we stepped the pulsewidth (at the same combline frequency) rather than stepping the optical frequency. This allowed us to perform a quasi-simultaneous characterization of noise across a large array of pulsewidths. The laser output was digitized at 3.7 GS/sec, and the ortho-RIN\cite{biedermann2009rin} was measured for each pulsewidth. The long off-time between pulses minimized cross-talk between neighboring pulses that would confound the measurements.

\subsubsection{Investigating the relationship between $t_{\textrm{OFF}}$ and $t_{\textrm{FP}}$}

To characterize how the duration of the programmed off-time between successive output pulses affects laser noise, we programmed the PCML laser to ping-pong between two output comblines (at 1559 and 1560 nm) repeatedly, with slowly increasing off-times between the pulses. The off times increased from 0.134 to 13.4 ns, except for the Finesse = 8000 configuration, which was 10$\times$ longer, as in the experiment $t_{\textrm{PW}}$. The pulsewidth remained fixed at a value in the region of stability as defined by the $t_{\textrm{PW}}$ experiment above. As before, ortho-RIN was measured for each of the four FP etalon configurations. For each etalon, the fixed pulsewidth was selected to provide low-noise operation as defined by the results of the previous section. 

\begin{figure}[]
\centering\includegraphics[width=8.2cm]{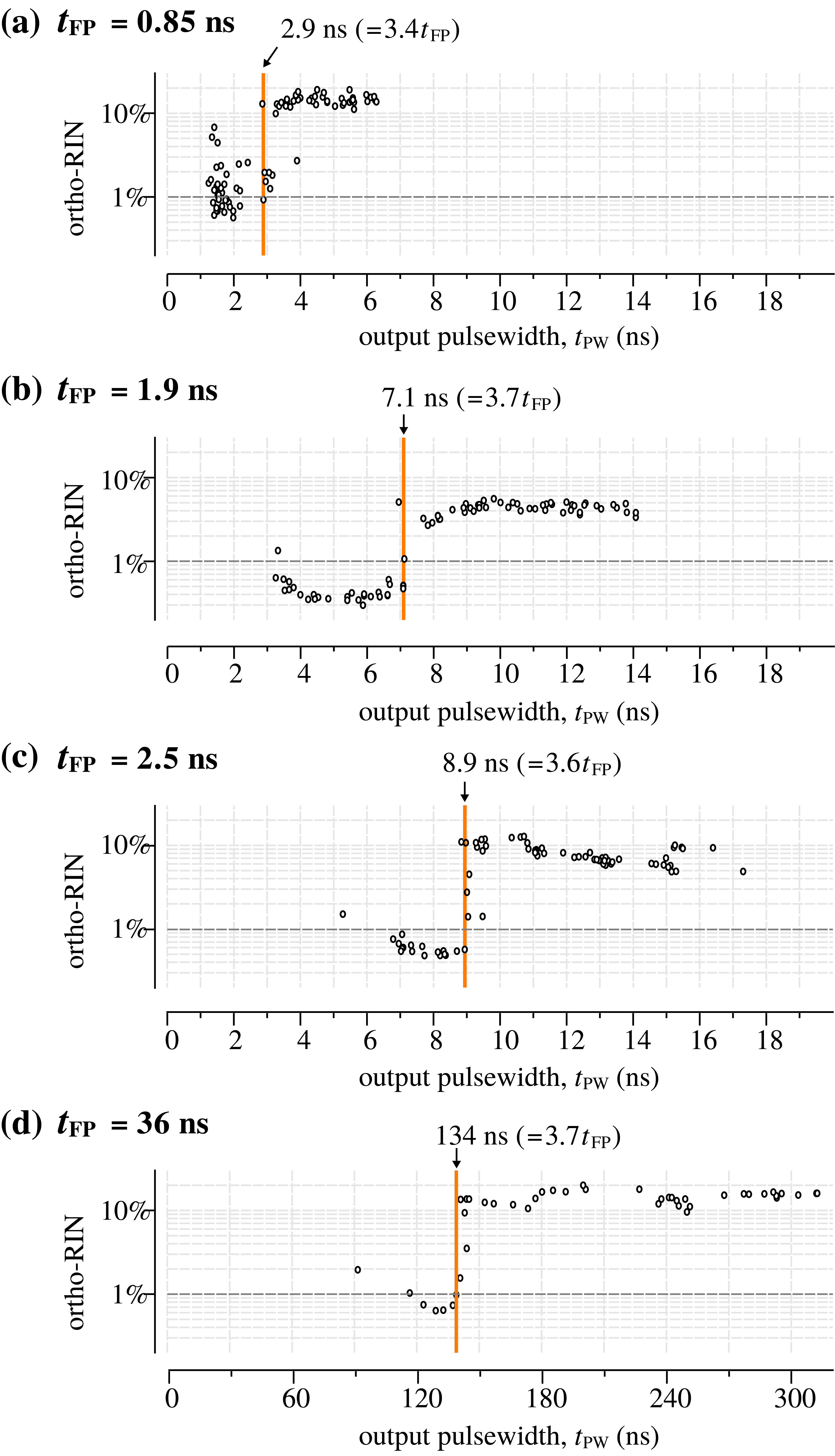}
\caption{Ortho-RIN when changing the output pulsewidth, $t_{\textrm{PW}}$. The plots correspond to (a) 150 [single pass], (b) 150 [double pass], (c) 500 [single pass], (d) 8000 [single pass] Finesse at 1560 nm. The free spectral ranges of each configuration ranges from 80-85 GHz. The orange lines indicate the inverse of the etalon linewidth.}
\label{fig:singlepulse}
\end{figure}

\section{Results}
\subsection{Noise Regimes based on $t_{\textrm{PW}}$ and $t_{\textrm{FP}}$}

Figure~\ref{fig:singlepulse} presents the ortho-RIN as a function of the measured pulsewidth for each etalon configuration. The achieved and programmed pulsewidths were highly, but not perfectly, correlated, and here we display the ortho-RIN as a function of the measured pulsewidths. The curves look similar when plotted as a function of programmed pulsewidths. For each etalon, we observed three operating regimes, detailed here.

\subsubsection{Pulsewidth $<<$ $t_{\textrm{FP}}$}
For commanded pulsewidths below the inverse of the etalon linewidth, it was not possible to achieve lasing. This is to be expected, as the optical bandwidth of these pulses exceeds the passband of the etalon, resulting in a large cavity loss that precludes lasing. 

\subsubsection{Pulsewidth $\approx$ $t_{\textrm{FP}}$}
As the commanded pulsewidth exceeds the inverse of the etalon linewidth, lasing is initiated. Here and up to approaching 3.5$\times$ the inverse of the etalon linewidth, the ortho-RIN is very low and a region of stability is observed. Note that the ortho-RIN values achieved in this regime are comparable to or better than most reported OCT systems.   

Note that the higher variability that is seen for the shorter pulsewidths (Fig.~\ref{fig:singlepulse}(a)) is likely due to the limitations of the sampling frequency of our RF signal generator (8 GSPS). For this etalon, the inverse of linewidth is approximately 3 ns, which implies that only 24 DAC samples are contained within the etalon "lifetime". We expect that if higher RF bandwidth drive signals were used, the variability in the measurements of Fig~\ref{fig:singlepulse}(a) would reduce significantly. Nonetheless, there are specific configurations for this etalon for which a stable ortho-RIN below 1\% can be achieved.

\subsubsection{Pulsewidth $>>$ $t_{\textrm{FP}}$}
For pulsewidths beyond $\approx$ 3.5$\times$ the inverse of the etalon linewidth, the laser shifts, almost discretely, to a higher noise operation. This behavior is observed for all four etalons, highlighting the value of matching the PCML pulsewidth to the etalon linewidth. 

\subsection{Noise Regimes based on $t_{\textrm{OFF}}$ and $t_{\textrm{FP}}$}

\begin{figure}[]
\centering\includegraphics[width=8.2cm]{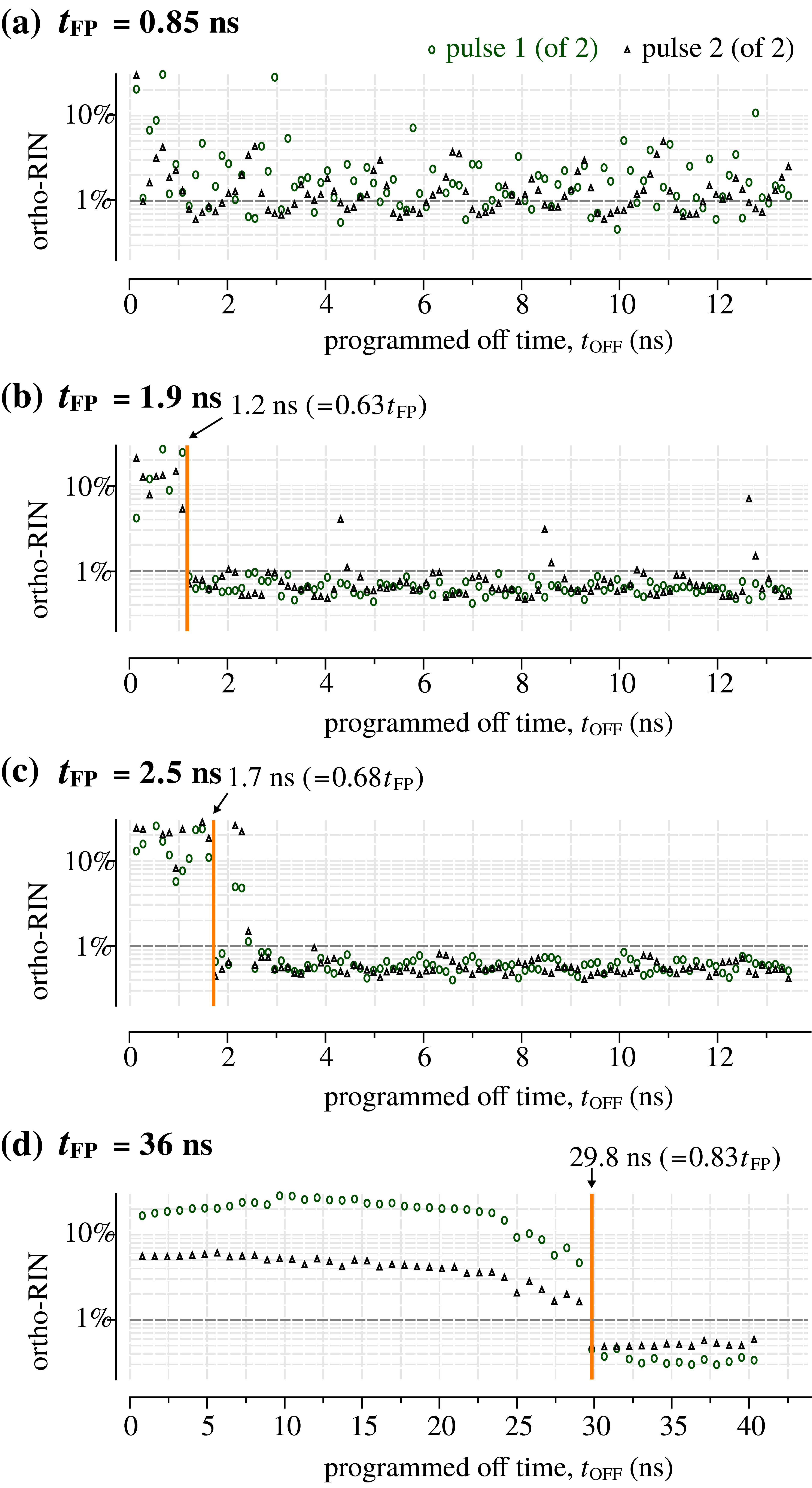}
\caption{Ortho-RIN when changing the programmed off-time, $t_{\textrm{OFF}}$. The plots correspond to (a) 150 [single pass], (b) 150 [double pass], (c) 500 [single pass], (d) 8000 [single pass] Finesse at 1560 nm. The free spectral ranges of each configuration ranges from 80-85 GHz. The orange lines indicate the empirical boundary between low-noise and high-noise operating regimes.}
\label{fig:doublepulse}
\end{figure}

Figure~\ref{fig:doublepulse} shows the performance of the PCML laser with respect to the off-time, $t_{\textrm{OFF}}$. Similar to the results seen in the pulsewidth experiment, low-noise and high-noise regimes are observed, separated by a narrow transition region. For the smallest linewidth/shortest off-times (Finesse of 150), it is difficult to perceive this boundary (i.e., to see the high-noise regime for short off-times), which is again likely a consequence of the finite digital-to-analog conversion sampling rate of our signal generators. For etalon finesses of 300, 500, and 8000, there is a clear boundary between unstable/high-noise and stable/low-noise operation, with a threshold time that is approximately $\frac{1}{4}$ to $\frac{1}{2}$ of the inverse of the etalon linewidth. Any $t_{\textrm{OFF}}$ beyond this limit produces low-noise lasing.

\subsection{Circular-ranging OCT with low-noise PCML}

With the timing parameters for optimal noise performance now known, OCT imaging was performed using the low-noise PCML source configured with the 500 Finesse (single-pass) FP etalon as a proof-of-principle experiment. 

\subsubsection{Laser performance}

Based on the results in Figs. \ref{fig:singlepulse} and \ref{fig:doublepulse}, the pulsewidth was set to 3.6 ns and the off-time 4.4 ns for OCT imaging. The PCML laser was programmed to step monotonically across 106 comblines starting at the shortest wavelength (1515 nm). The total optical bandwidth of the source was 72 nm (= 106 x 85 GHz etalon FSR). The pulse repetition time was 8 ns (= 3.6 ns + 4.4 ns), making the A scan rate equal to 1.179 MHz. 

\begin{figure}[]
\centering\includegraphics[width=8.0cm]{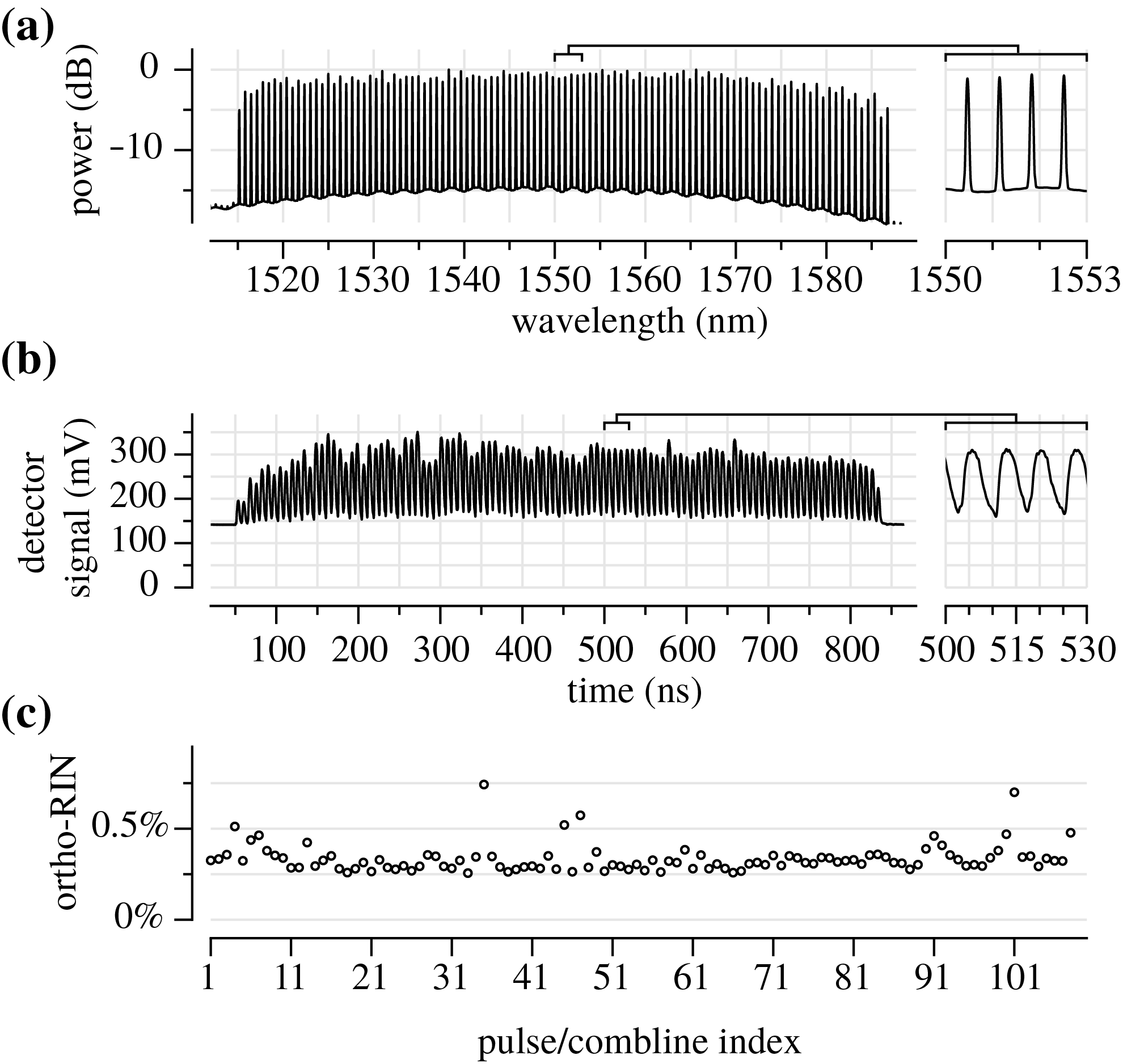}
\caption{(a) OSA spectrum of the PCML laser. (b) Time trace and (c) ortho-RIN at the peak of each pulse.}
\label{fig:explaser}
\end{figure}

The laser output spectrum and power trace are shown in Fig.~\ref{fig:explaser}(a) and Fig.~\ref{fig:explaser}(b), respectively. The ortho-RIN of each combline, shown in Fig.~\ref{fig:explaser}(c), was measured to be below 1\% throughout the spectrum, with a mean ortho-RIN of 0.34 \% and a maximum ortho-RIN of 0.74 \%. Note that both the spectral trace and the temporal trace show some structure in the PCML pulse output powers, but this is a fixed-pattern (repeatable) structure and not indicative of instability or noise. 

\subsubsection{Optical coherence tomography imaging}

We used a simple Mach-Zehnder interferometer with a previously described passive optical quadrature demodulation circuit \cite{siddiqui2015compensation} to perform circular-ranging OCT using the low-noise PCML laser as the input light source. The microscope and detection system were identical to that described in the prior demonstration of PCML \cite{kim2020stepped}. A booster SOA was used to amplify the output power of the PCML such that the power on the sample reached 35 mW. At this power, the system sensitivity was measured to be 103 dB. The 6 dB roll-off depth was 7.4 cm with the mirror translation in double-pass configuration, corresponding to 14.8 cm of optical delay. It is expected that this roll-off is limited by the presence of the booster amplifier, and further optimization of roll-off would be possible in the future.

\begin{figure}[]
\centering\includegraphics[width=8.0cm]{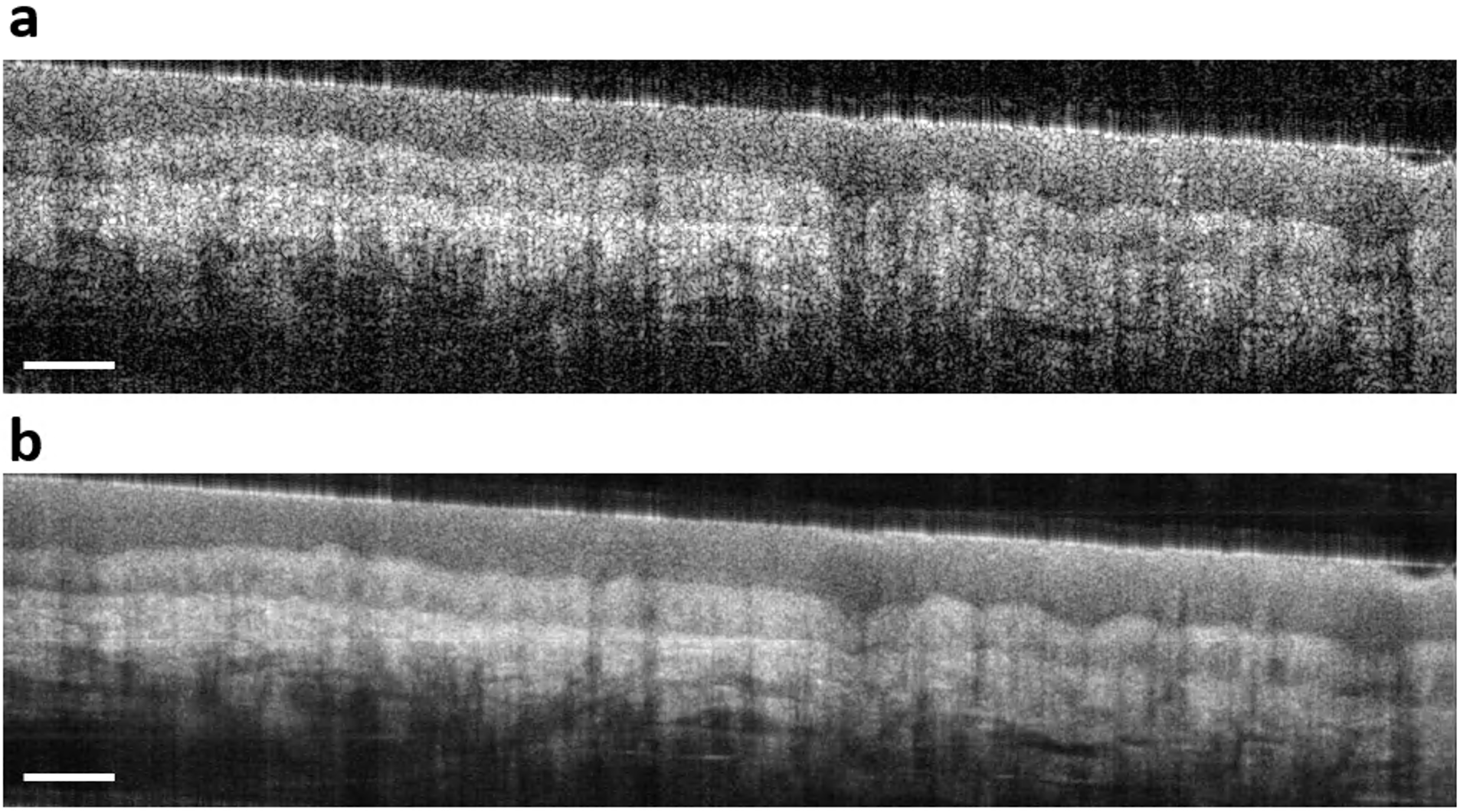}
\caption{(a) Non-averaged cross sectional image of the skin of the finger acquired from the PCML-OCT system. (b) The finger image at the same location following averaging of 25 consecutive B-scans. The height of the cross sections correspond to the circular range of the system. Dynamic range: 38 dB. Scale bar: 500 $\mu$m.}
\label{fig:finger}
\end{figure}

The system was used to image the skin of a finger, the results of which can be found in Fig. \ref{fig:finger}. Figure \ref{fig:finger}(a) shows the original non-averaged cross-sectional image, which shows a 20 dB SNR improvement relative to the first-generation PCML laser \cite{kim2020stepped}. An out-of-plane-averaged cross-section can be found in Fig. \ref{fig:finger}(b), visually eliminating intensity fluctuation due to speckle \cite{wu2013assessment}. These images are on par with those achieved from traditional swept-source OCT systems.

\section{Discussion}

We have demonstrated an operating regime, defined by the temporal properties of the output pulse stream relative to the inverse of the etalon linewidth, for which the PCML laser exhibits low-intensity noise. This adds low noise to the list of advantages of the PCML laser which include support for arbitrary output sequences and intrinsic locking of the output to electronic drive signals and thereby to acquisition electronics. In this work, the sweep rate is limited by the RF signal drive bandwidth, but the laser architecture and electro-optic modulators can support higher-speed operation should a faster PCML laser be desired. Like the prior description, this work describes PCML lasing at 1550 nm wavelengths. Current work is underway to develop PCML lasers at the more common OCT wavelengths of 1300 nm and 1050 nm now that stability and low noise have been achieved.

This work shows the noise advantage to operating the PCML with distinct output pulses. The PCML laser loses some of its generalizability by constraining the pulsewidth to a range set by the (fixed) etalon properties, but it is possible to generate an effectively longer pulse at a given combline by generating multiple short pulses at that combline. After low-pass filtering by the OCT detection system, this provides nominally the same optical signal as would be generated by a single long pulse. 

The PCML should be operated to create distinct output pulses with a well-defined off-period between the pulses. Nonetheless, this off-time can be relatively short, allowing a high duty cycle. If we take the double-passing of the 150 Finesse etalon as an example, the optimal pulsewidth range is approximately 5-7 ns, while the required off-time is only ~1 ns. Of note, a recent work shows that there are intrinsic CR-specific SNR advantages to using a stepped frequency comb source with strong amplitude pulsation in CR \cite{lippok2022rf}, which aligns fortuitously to the low-noise operating regimes of the PCML laser. 

\section{Conclusion}

In this work, we have demonstrated that operating the PCML laser with pulse widths and pulse separations tuned to match the characteristic time response of the intracavity FP etalon can reduce output noise by more than an order of magnitude. This significant noise reduction translates into a marked improvement in OCT imaging performance, positioning PCML as a competitive light source for circular-ranging OCT.

\textbf{Funding}
The research reported in this publication was supported in part by the Center for Biomedical OCT Research and Translation through Grant Number P41EB015903.

\bibliographystyle{unsrtnat}
\bibliography{sample}

\begin{thebibliography}{14}
\providecommand{\natexlab}[1]{#1}
\providecommand{\url}[1]{\texttt{#1}}
\expandafter\ifx\csname urlstyle\endcsname\relax
  \providecommand{\doi}[1]{doi: #1}\else
  \providecommand{\doi}{doi: \begingroup \urlstyle{rm}\Url}\fi

\bibitem[Izatt and Choma(2008)]{izatt2008theory}
Joseph~A Izatt and Michael~A Choma.
\newblock Theory of optical coherence tomography.
\newblock In \emph{Optical Coherence Tomography: Technology and Applications}, pages 47--72. Springer, 2008.

\bibitem[Siddiqui and Vakoc(2012)]{siddiqui2012optical}
Meena Siddiqui and Benjamin~J Vakoc.
\newblock Optical-domain subsampling for data efficient depth ranging in {F}ourier-domain optical coherence tomography.
\newblock \emph{Optics Express}, 20\penalty0 (16):\penalty0 17938--17951, 2012.

\bibitem[Siddiqui et~al.(2018)Siddiqui, Nam, Tozburun, Lippok, Blatter, and Vakoc]{siddiqui2018high}
Meena Siddiqui, Ahhyun~S Nam, Serhat Tozburun, Norman Lippok, Cedric Blatter, and Benjamin~J Vakoc.
\newblock High-speed optical coherence tomography by circular interferometric ranging.
\newblock \emph{Nature Photonics}, 12\penalty0 (2):\penalty0 111--116, 2018.

\bibitem[Bouma et~al.(2022)Bouma, de~Boer, Huang, Jang, Yonetsu, Leggett, Leitgeb, Sampson, Suter, Vakoc, Villiger, and Wojtkowski]{bouma2022primer}
Brett~E. Bouma, Johannes~F. de~Boer, David Huang, Ik-Kyung Jang, Taishi Yonetsu, Cadman~L. Leggett, Rainer Leitgeb, David~D. Sampson, Melissa Suter, Ben~J. Vakoc, Martin Villiger, and Maciej Wojtkowski.
\newblock Optical coherence tomography.
\newblock \emph{Nature Reviews Methods Primers}, 2\penalty0 (1):\penalty0 79, Oct 2022.
\newblock ISSN 2662-8449.
\newblock \doi{10.1038/s43586-022-00162-2}.
\newblock URL \url{https://doi.org/10.1038/s43586-022-00162-2}.

\bibitem[Tozburun et~al.(2014)Tozburun, Siddiqui, and Vakoc]{tozburun2014rapid}
Serhat Tozburun, Meena Siddiqui, and Benjamin~J Vakoc.
\newblock A rapid, dispersion-based wavelength-stepped and wavelength-swept laser for optical coherence tomography.
\newblock \emph{Optics Express}, 22\penalty0 (3):\penalty0 3414--3424, 2014.

\bibitem[Lippok et~al.(2019)Lippok, Siddiqui, Vakoc, and Bouma]{lippok2019extended}
Norman Lippok, Meena Siddiqui, Benjamin~J Vakoc, and Brett~E Bouma.
\newblock Extended coherence length and depth ranging using a {F}ourier-domain mode-locked frequency comb and circular interferometric ranging.
\newblock \emph{Physical Review Applied}, 11\penalty0 (1):\penalty0 014018, 2019.

\bibitem[Khazaeinezhad et~al.(2017)Khazaeinezhad, Siddiqui, and Vakoc]{khazaeinezhad2017spml}
Reza Khazaeinezhad, Meena Siddiqui, and Benjamin~J. Vakoc.
\newblock 16\&\#x2009;\&\#x2009;mhz wavelength-swept and wavelength-stepped laser architectures based on stretched-pulse active mode locking with a single continuously chirped fiber bragg grating.
\newblock \emph{Opt. Lett.}, 42\penalty0 (10):\penalty0 2046--2049, May 2017.
\newblock \doi{10.1364/OL.42.002046}.
\newblock URL \url{https://opg.optica.org/ol/abstract.cfm?URI=ol-42-10-2046}.

\bibitem[Kim and Vakoc(2020)]{kim2020stepped}
Tae~Shik Kim and Benjamin~J Vakoc.
\newblock Stepped frequency comb generation based on electro-optic phase-code mode-locking for moderate-speed circular-ranging {OCT}.
\newblock \emph{Biomedical Optics Express}, 11\penalty0 (7):\penalty0 3534--3542, 2020.

\bibitem[Huber et~al.(2006)Huber, Wojtkowski, and Fujimoto]{huber2006fdml}
R.~Huber, M.~Wojtkowski, and J.~G. Fujimoto.
\newblock Fourier domain mode locking (fdml): A new laser operating regime and applications for optical coherence tomography.
\newblock \emph{Opt. Express}, 14\penalty0 (8):\penalty0 3225--3237, Apr 2006.
\newblock \doi{10.1364/OE.14.003225}.
\newblock URL \url{https://opg.optica.org/oe/abstract.cfm?URI=oe-14-8-3225}.

\bibitem[Vaughan(2017)]{vaughan2017fabry}
M~Vaughan.
\newblock \emph{The {Fabry-Perot} interferometer: history, theory, practice and applications}, chapter~6, pages 1--40.
\newblock Routledge, 2017.

\bibitem[Biedermann et~al.(2009)Biedermann, Wieser, Eigenwillig, Klein, and Huber]{biedermann2009rin}
Benjamin~R. Biedermann, Wolfgang Wieser, Christoph~M. Eigenwillig, Thomas Klein, and Robert Huber.
\newblock Dispersion, coherence and noise of fourier domain mode locked lasers.
\newblock \emph{Opt. Express}, 17\penalty0 (12):\penalty0 9947--9961, Jun 2009.
\newblock \doi{10.1364/OE.17.009947}.
\newblock URL \url{https://opg.optica.org/oe/abstract.cfm?URI=oe-17-12-9947}.

\bibitem[Siddiqui et~al.(2015)Siddiqui, Tozburun, Zhang, and Vakoc]{siddiqui2015compensation}
Meena Siddiqui, Serhat Tozburun, Ellen~Ziyi Zhang, and Benjamin~J Vakoc.
\newblock Compensation of spectral and {RF} errors in swept-source {OCT} for high extinction complex demodulation.
\newblock \emph{Optics Express}, 23\penalty0 (5):\penalty0 5508--5520, 2015.

\bibitem[Wu et~al.(2013)Wu, Tan, Pappuru, Duan, and Huang]{wu2013assessment}
Wei Wu, Ou~Tan, Rajeev~R Pappuru, Huilong Duan, and David Huang.
\newblock Assessment of frame-averaging algorithms in {OCT} image analysis.
\newblock \emph{Ophthalmic Surgery, Lasers and Imaging Retina}, 44\penalty0 (2):\penalty0 168--175, 2013.

\bibitem[Lippok and Vakoc(2022)]{lippok2022rf}
Norman Lippok and Benjamin~J Vakoc.
\newblock {RF} properties of circular-ranging {OCT} signals.
\newblock \emph{Optics Letters}, 47\penalty0 (7):\penalty0 1903--1906, 2022.

\end{thebibliography}

\end{document}